\documentclass[]{mn2e}

\usepackage{graphicx}
\usepackage[dvips]{color}

\title[Fast evolving size of early-type galaxies at $z>2$]{Fast evolving size of early-type galaxies at $z>2$ and the 
role of 
dissipationless ({\it dry}) merging} \author[A. Cimatti, C. Nipoti, P. Cassata]{A. Cimatti$^{1}$\thanks{E-mail:
a.cimatti@unibo.it}, 
C. Nipoti$^{1}$, 
P. Cassata$^{2}$\\
$^{1}$Universit\`a di Bologna, Dipartimento di Astronomia, Via Ranzani
1, I-40127, Bologna, Italy\\
$^{2}$Laboratoire d'Astrophysique de Marseille - LAM, Universit\'e
d'Aix-Marseille \& CNRS, UMR7326, 38 rue F. Joliot-Curie, \\F-13388
Marseille Cedex 13, France
}
\begin{document}

\date{Accepted 2012 xxx xxx. Received 2011 XX XX; in original form 2011 XX XX}

\pagerange{\pageref{firstpage}--\pageref{lastpage}} \pubyear{2002}

\maketitle

\label{firstpage}

\begin{abstract}
We present the analysis of a large sample of early-type galaxies
(ETGs) at $0<z<3$ aimed at tracing the cosmic evolution of their size
and compare it with a model of pure dissipationless ({\it dry})
merging in the $\Lambda$CDM framework. The effective radius $R_e$
depends on stellar mass ${\cal M}$ as $R_e({\cal M}) \propto {\cal
M}^{\alpha}$ with $\alpha \sim 0.5$ at all redshifts. The redshift
evolution of the mass- or SDSS-normalized size can be reproduced as
$\propto (1+z)^{\beta}$ with $\beta \sim -1$, with the most massive 
ETGs possibly showing the fastest evolutionary rate ($\beta \sim -1.4$). This 
size evolution slows down significantly to $\beta \sim -0.6$ if 
the ETGs at $z>2$ are removed from the sample, suggesting an accelerated
increase of the typical sizes at $z>2$, especially for the ETGs with
the largest masses. A pure {\it dry} merging $\Lambda$CDM model
is marginally consistent with the average size evolution at
$0<z<1.7$, but predicts descendants too compact for $z>2$ progenitor
ETGs. This opens the crucial question on what physical mechanism
can explain the accelerated evolution at $z>2$, or whether an unclear
observational bias is partly responsible for that.
\end{abstract}


\section{Introduction}

Early-type galaxies (ETGs) are important probes of structure formation
and massive galaxy evolution. At $0<z<1$, the ETG stellar mass
function shows a {\it downsizing} evolution apparently difficult to
reproduce with the current models of galaxy formation, with the
majority of massive ETGs (${\cal M} > 10^{11}$ M$_{\odot}$) already in
place at $z\approx0.7$, (Pozzetti et al. 2010 and references
therein). At $z>1$, the information is still incomplete, but {\it bona fide} ETGs have
been identified up to $z\sim 2.5$ (e.g.  Kriek et al. 2006; Cimatti et
al. 2008 and references therein). These high-$z$ ETGs are
characterized by old stars (1-3 Gyr), $e$-folding decaying star formation
timescales $\tau \sim$0.1--0.3 Gyr, low specific star formation rates
(SSFR$<10^{-2}$ Gyr$^{-1}$), low dust extinction, large stellar masses
(${\cal M} > 10^{11} M_{\odot}$), spheroidal morphologies (although some
of these systems have a disk-like component; van der Wel et al. 2011),
and number densities growing rapidly from $z> 3$ to $z \sim 1$ (e.g.
Fontana et al. 2009; Dom\'inguez-S\'anchez et al. 2011; Brammer et
al. 2011).

A puzzling property of ETGs at $z>1$ is that they have smaller
sizes, down to effective radii $R_e < 1$ kpc, and correspondingly
higher internal mass densities than present-day ETGs with the same
mass (e.g. Daddi et al. 2005; Trujillo et al. 2006; van der Wel et
al. 2008; Cimatti et al. 2008; Buitrago et al. 2008; Saracco, 
Longhetti \& Andreon 2009; Williams et al. 2010;
Cassata et al.  2011; Damjanov et al. 2011; Newman et al. 2011, and
references therein). It is not clear yet whether the environment plays
(e.g. Papovich et al. 2011; Cooper et al. 2011) or not (e.g. 
Rettura et al. 2010) a role in the ETG size evolution. 
The few available measurements of stellar
velocity dispersions are consistent with those expected from the ETG
sizes and confirm that these systems are truly massive (Cenarro et
al. 2009; Cappellari et al.  2009; Onodera et al. 2010; van Dokkum,
Kriek \& Franx 2009; van de Sande et al. 2011).  Several models have
been proposed to explain the size--mass evolution, including
dissipationless ({\it dry}) major and minor merging, adiabatic
expansion driven by stellar mass loss and/or strong feedback, and
smooth stellar accretion (e.g. Khochfar \& Silk 2006; Bournaud, 
Jog \& Combes 2007; Fan et
al. 2008; Nipoti et al. 2003, 2009a, 2009b; Naab et al. 2009; Hopkins
et al.  2009; Oser et al. 2011). However, the global picture is far
from being clear.
In this paper, we exploit a large sample of ETGs in order to investigate 
the evolution of their size as a function of redshift and mass, and
compare it with the predictions of cosmological models of structure
formation. 

\section{The sample}

In order to improve statistically on previous studies, we selected a
large sample of 1975 ETGs at $0.2<z<3$ by collecting data from the
literature and public data, requiring the availability of
spectroscopic redshifts (or high-quality photometric redshifts
for $z>1.4$), stellar masses, sizes ($R_e$) and, when possible, 
age of the stellar population (see Tab. 1). 
The ETGs in the different subsamples share the global property to 
have been originally selected based on the combination of colors, 
spectra (or sometimes also SSFR) typical of old/passive galaxies with the 
confirmation of spheroidal (E/S0) morphology, or vice versa. 
We recall that selection criteria for ETGs are strongly correlated, 
with up to $\sim$85\% color/spectra-selected ETGs being also 
morphologically E/S0 (e.g. Renzini 2006 and references therein). The ETG sizes were 
generally measured in the observed-frame red-optical for 
low-/intermediate-redshift samples and/or in the near-infrared for 
higher redshifts, i.e. typically sampling the rest-frame optical 
region at all redshifts. Recently, Damjanov et al. (2011) and 
Cassata et al. (2011) have shown that the sizes measured in the 
rest-frame UV and in the optical correlate very strongly with 
each other, thus excluding substantial biases dependent on 
the wavelength at which the size was measured. The SDSS sample of 
Hyde \& Bernardi (2009) was included as the reference sample at 
$z\sim 0$.

\begin{table}
\caption{ETG subsamples}
\begin{tabular}{l l l l l l}
\hline
Sample & N & Redshift & Age & Ref. \\ 
       &   &          &     &      \\ 
\hline
SDSS & 59500 & $0<z<0.4$ & yes & 1 \\
COSMOS/zCOSMOS & 950 & $0<z<1$ & no & 2 \\
GOODS-N+S & 469 & $0<z^*<2$ & yes & 3 \\
Literature& 465 & $0.2<z<2.7$ & no & 4 \\
GMASS & 45 & $1.4<z^*<3$ & no & 5 \\
COSMOS & 12 & $1.4<z^*<1.8$ & yes & 6 \\
XMMU J2235-2557 & 11 & $z$=1.39 & yes & 7 \\
K20-0055 & 9 & $0.7<z<1.2$ & yes & 8 \\
POWIR & 6 & $1.2<z^*<1.8$ & no & 9 \\
K20 & 4 & $1.6<z<1.9$ & yes & 10 \\
1255--0 & 1 & $z=2.186$ & yes & 11 \\
FW-4871 & 1 & $z=1.902$ & yes & 12 \\
\hline
\end{tabular}
\footnotesize{$z^*$: a fraction of redshifts is photometric, 
Age: available stellar ages. 1. Hyde \& Bernardi 2009, 
2. http://cosmos.astro.caltech.edu/data/index.html, Scarlata et al. 2007, 
Moresco et al. 2010, 3. Cassata et al. 2011, 4. Damjanov
et al. 2011, 5. Cassata al. 2008, 6. Mancini et al. 2010,
7. Strazzullo et al. 2010, 8. di Serego Alighieri et al. 2005,
9. Carrasco, Conselice \& Trujillo 2010 ($n_{Sersic}>2$), 10. Cimatti et al. 2004, 11. 
van Dokkum et al. 2009, 12. van Dokkum \& Brammer 2010.
}
\end{table}

The different subsamples were harmonized to the same cosmology
($H_0=70$ km s$^{-1}$ Mpc$^{-1}$, $\Omega_{\rm m}= 0.3$, 
$\Omega_{\Lambda}=0.7$), and the stellar masses and ages rescaled to 
the Maraston (2005) stellar population synthesis models with the 
Chabrier Initial Mass Function (IMF) by using empirical scaling 
relations of Pforr et al. (2012).

Based on the information available in the literature, the stellar 
mass completeness of most subsamples is $\log {\cal M}_{comp}/ 
M_{\odot}=10.5$ at all redshifts. However, a few subsamples have
$\log {\cal M}_{comp}/ M_{\odot} \sim 10.8-11$: GN/DEIMOS (from 
Damjanov et al. 2011) and zCOSMOS at $z<1.3$, MUNICS, K20, Mancini 
et al. (2010) and van Dokkum et al. (2008) at $1.3<z<2$, Cassata et al. 
(2011) and van Dokkum et al. (2008) at $z>2$. The final sample was
reduced to 1080 galaxies in order to avoid mass incompleteness 
effects (see Section 4 for details).

\section{The size -- mass relation}

Figure 1 shows the size--mass relation in three redshift ranges with 
a comparable number of galaxy in each bin. A power-law fit,
$R_e \propto {\cal M}^{\alpha}$, applied to non-SDSS ETGs with 
$\log {\cal M}/ M_{\odot}>$10.5 provides $\alpha=0.52\pm0.05, 0.47\pm0.04, 
0.50\pm0.04$ from low to high redshift in the three bins of Fig. 1,
with $\alpha$ basically independent of redshift. For instance, the 
ETGs with $z>2$ have $\alpha=0.45\pm0.11$. This result is consistent 
with recent works (Damjanov et al. 2011; Newman et al. 2011), and does 
not depend significantly on the choice of the redshift bin limits or 
the minimum stellar mass cuts. 
In comparison, the SDSS ETGs with $\log {\cal M}/ M_{\odot}>$ 10.5 have 
$\alpha=$0.58$\pm0.01$, consistent with Shen et al. (2003). 

\begin{figure}
%\begin{figure*}
\begin{center}
\includegraphics[width=0.98\linewidth]{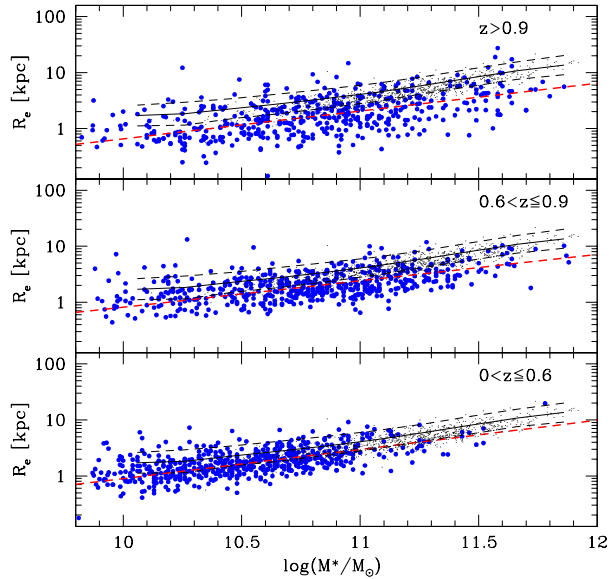}
\caption{The size -- stellar mass relation for $\log {\cal M}/
M_{\odot}>$10.5. Black and blue symbols indicate SDSS and 
non-SDSS ETGs respectively. Only SDSS ETGs older than 10 Gyr 
are shown in order to avoid excessive crowding. The black solid line 
shows the size--mass relation of Shen et al. (2003) and its
$\pm 1 \sigma$ scatter (dashed black lines). Red dashed lines:
best fit relations using $R_e \propto {\cal M}^{\alpha}$.
}
\label{test figure}
\end{center}
%\end{figure*}
\end{figure}

\section{The size -- redshift relation}

Fig. 2 shows the redshift evolution of the ETG size in two
complementary ways: the mass-normalized radius ($R_e(z)/{\cal 
M}_{11}^{\alpha}$, where ${\cal M}_{11}={\cal M}/10^{11}$ M$_{\odot}$, 
adopting $\alpha=0.55$ as representative value), and radius 
normalized to the average size of SDSS ETGs ($R_e(z)/R_e(SDSS)$)
in three mass bins. 
Both quantities are useful to derive the size evolution 
independently of the correlation between $R_e$ and ${\cal M}$.
The evolution is parametrized by the usual functional form $size
\propto (1+z)^{\beta}$. Clearly, this parametrization does not 
mean that {\it all} high-$z$ ETGs are the direct progenitors 
of {\it all} low-$z$ ETGs because these galaxies evolve in 
the redshift range $0<z<3$ through several processes and 
increase their number density and mass, but it has simply the 
statistical meaning of indicating how the {\it typical}
sizes compare at different redshifts.

In order to mitigate the potential effects of stellar mass 
incompleteness, for each mass bin of Fig.2 (top three panels),
the galaxies with ${\cal M}<{\cal M}_{comp}(z)$ were removed
from each subsample. Thus, each mass bin is always complete down to
the minimum mass of the bin. For instance, the ETGs with 
${\cal M}<10^{11}$ M$_{\odot}$ at $z \geq 1$ and ${\cal M}<
10^{10.6}$ M$_{\odot}$ at 
$z \geq 2$ have been excluded from the zCOSMOS and the Cassata
et al. (2011) sample respectively.

For the mass-normalized radius (Fig.2, bottom panel), we derive 
$\beta$= -1.06 $\pm 0.14$(-1.24 $\pm 0.15)$ for $\log {\cal M}/ 
M_{\odot}>10.5(10.9)$, in agreement with the literature 
(Damjanov et al. 2011, Newman et al. 2011).
This result is stable against changing $\alpha$ between 0.4 and 0.7,
the boundaries of the redshift bins, and the minimum stellar mass. 
The consistency of the $R_e \propto {\cal M}^{\alpha}$ and
$(R_e/{\cal M}_{11}^{\alpha}) \propto (1+z)^{\beta}$ relations with 
previous results confirms that our sample, despite its somehow 
heterogeneous composition, can be reliably used to perform
further detailed studies.

\begin{figure}
%\begin{figure*}
%\begin{center}
\includegraphics[width=0.98\linewidth]{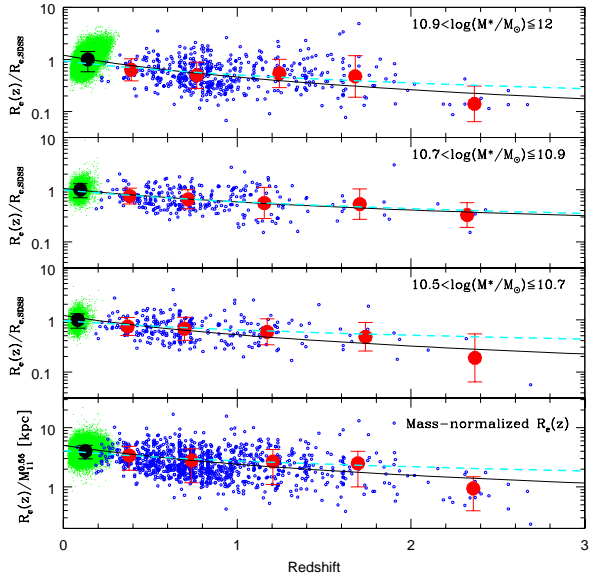}
\caption{{\it Bottom panel:} the mass-normalized radius 
($R_e/{\cal M}_{11}^{0.55}$) of ETGs with $\log {\cal M}/
M_{\odot}>$10.5 as a function of $z$. {\it Top three panels:} 
the fractional radius (relative to SDSS; $\langle R_e(SDSS) \rangle$=
2.51, 3.10, 5.61 kpc). In all panels, green and blue small symbols 
indicate individual SDSS and non-SDSS ETGs respectively, 
whereas the large black and red filled circles are the average 
values ($\langle R_e/R_e(SDSS)\rangle$) of SDSS and non-SDSS 
respectively, each with the $\pm 1\sigma$ scatter. The black solid 
line is the best fitting function to the red and black points. 
The redshift bins adopted for the non-SDSS ETGs in this figure are 
$0<z<0.5$, $0.5<z<1$, $1<z<1.5$, $1.5<z<2$, $2<z<3$. The
cyan dashed lines show the best fits for the case of ETGs with $z<2$
(redshift bins: $0<z<0.3$, $0.3<z<0.6$, $0.6<z<1$, $1<z<1.5$,
$1.5<z<2$).
}
%\label{test figure}
%\end{center}
%\end{figure*}
\end{figure}

Fig. 2 (top three panels) shows $R_e$ normalized to $R_e(SDSS)$.
For the three bins of increasing mass, we derive $\beta$=-1.23$\pm$
0.15, -0.87$\pm$0.14 and -1.39$\pm$0.13. This suggests that $\beta$
does not significantly depend on stellar mass, although a steepening
is suggested in the most massive bin. 
These results are stable, within the statistical uncertainties, against 
changes in the minimum stellar mass in the lowest mass bin, in the 
boundaries of the mass bins, and in the minimum age of the SDSS ETGs. 

Fig. 2 also shows that the ETG size growth rate seems to
increase significantly at high redshift, suggesting a slower 
evolution at $z<2$. For the mass-normalized radius (bottom panel),  
$\beta$ becomes -0.57$\pm$0.15 if the ETGs at $z_{cut}>2$ are excluded. 
Similarly for $R_e/R_e(SDSS)$, $\beta$ becomes  $\beta$=-0.73$\pm$0.14, 
-0.76$\pm$0.13 and -0.75$\pm$ 0.14 if ETGs at $z>2$ are excluded
in the three bins of increasing mass. We note that the flattening is 
particularly pronounced and significant in the bin of largest masses. 
The flattening of $\beta$ is also present, within the statistical 
uncertainties, against changing the limits of the redshift bins if 
we cut ETGs around $1.5<z_{cut}<2$. 
For instance, if the ETGs with $z_{cut}>1.5(1.7)$ 
are excluded in the bin with $\log {\cal M}/ M_{\odot}>$10.9, we 
derive $\beta$=-0.95(-0.72)$\pm$0.15. The overall results presented
in this section do not change if median values are used instead 
of the average ones, nor if the SDSS data points are excluded. 

The potential role of the so-called progenitor bias (e.g. Saglia et
al. 2010 and references therein) has been assessed through an {\it age
filtering} by comparing the size of ETGs having ages compatible with 
the cosmic time passed from high-$z$ to low-$z$. For a given redshift 
range $z_1<z<z_2$ and
average redshift $\bar z$, we estimated $\beta$ by comparing the
average size of the ETGs (having an average age 
$t_{age}(\bar z) \pm \sigma_{t(z)}$) with the size of SDSS ETGs with
ages at $z_0$ between $[t_{age}(\bar z)-\sigma_{t(z)}] +\tau(z-z_0)$
and $[t_{age}(\bar z)+\sigma_{t(z)}]+\tau(z-z_0)$, where $\tau(z-z_0)$
is the cosmic time passed from $z$ to $z_0$.  Based on this approach,
no significant or systematic changes of $\beta$ have been found for a
variety of tested redshift ranges.  For example, for ETGs with $\log {\cal
M}/ M_{\odot}>$10.9 at $0.7<z<1.3$, we derive $\beta= -1.06\pm 0.2$
and $\beta=-0.97\pm 0.2$ respectively with and without applying the
above ¨age filtering¨. Similarly, if we select the 
most distant ETGs ($1.8<z<3$), we obtain $\beta= -1.29\pm 0.2$ 
and $\beta=-1.23\pm0.2$. 
If we take these results at face value, this implies that most
of the $R_e(z)$ evolution is unlikely to be the result of a progenitor
bias due to high-$z$ ETGs being preferentially selected to be redder
and more compact than lower-$z$ younger and larger ETGs missed at
high-$z$. However, we recall that, due to the heterogeneous estimates 
of the stellar ages in our sample, this result should be confirmed
with larger and more homogeneous samples. On the other hand, we note 
that large ETGs are indeed absent at $1.4<z<3$ in the GMASS subsample 
(Fig. 2), that is selected based solely on morphology 
irrespective of colours (Cassata et al. 2008).

\section{Comparison with a $\Lambda$CDM model}

Dissipationless ({\it dry}) merging is one of the few mechanisms 
known to make galaxies
less compact (e.g., Nipoti et al. 2003; Naab et al. 2009), so it is
often invoked to explain the observed size evolution of ETGs.  Here we
briefly address the question of whether the observed size evolution is
consistent with the merger histories of concordance $\Lambda$ cold
dark matter ($\Lambda$CDM) cosmology. For this purpose, we use the
$\Lambda$CDM-based merger models presented in Nipoti et al. (2012,
hereafter N12), which are such that the variation in surface-mass
density is maximized, because dissipative effects are neglected. In
particular, we adopt here model B of N12, which is the most strongly
evolving of their models, so the predicted size evolution must be
considered an upper limit. Here we briefly describe the main
properties of the model, but we refer the reader to N12 for details.
For an observed ETG with measured stellar mass ${\cal M}$ and
effective radius $R_e$, the model allows to calculate the redshift
evolution of ${\cal M}$ and $R_e$ via analytic functions calibrated on
$N$-body simulations. The galaxy is first assigned a halo mass using
the redshift-dependent stellar to halo mass relation of Behroozi,
Conroy, \& Wechsler (2010). The halo growth-history is then computed
using Fakhouri et al. (2010) fit to halo merger histories in the
Millenium I and II simulations (Springel et al. 2005; Boylan-Kolchin
et al. 2009). The associated growth of ${\cal M}$ is obtained by
assigning stellar mass to satellite halos with Behroozi et al. (2010)
recipe, and considering only mergers with mass ratio $>$0.03 (to exclude 
cases with too long merging time). Finally, the corresponding variation in
$R_e$ is computed using analytic functions verified with $N$-body
simulations of minor and major dry mergers between spheroids.
For observed high-$z$ ETGs, we compute the predicted ${\cal M}(z)$ and
$R_e(z)$ up to $z=0.14$, which is the average redshift of massive
($\log {\cal M}/ M_{\odot}>10.9$) SDSS ETGs. The evolution in the
${\cal M}$-$R_e$ plane is shown in Fig. 3, taking as progenitors the
observed ETGs with $2<z<3$ ($\langle z\rangle\simeq 2.4$) and those
with $1.5<z<2$ ($\langle z\rangle\simeq 1.7$).  In both cases the
present-day descendants tend to be more compact than real $z\sim 0$
ETGs, but the deviation from the SDSS ${\cal M}$-$R_e$ relation is
larger than the observed scatter ( $\sim 0.15$ $\log R_e$ at given
${\cal M}$) only when $z\sim2.4$ progenitors are considered: the
$z\sim0$ descendants have median vertical offset from the SDSS
best-fit $\Delta \log R_e\simeq-0.3$(-0.1) for $z\sim2.4$(1.7)
progenitors.  In the case of $z\sim 2.4$ progenitors, also the $z=1$
model descendants are more compact than real $z\sim 1$ ETGs.  We
conclude that the $z>2$ ETGs are so compact that, even according to
extreme pure {\it dry}-merger models, their low-$z$ descendants are
predicted to be significantly more compact than present-day ETGs. On
the other hand, the milder size evolution observed since $z\sim1.7$ is
marginally consistent with $\Lambda$CDM {\it dry}-merger models,
though the model descendants are distributed in the ${\cal M}$-$R_e$
plane with larger scatter than the observed ETGs (see also N12). An
additional problem is the presence, among the predicted $z\sim 0$
descendants, of outliers, i.e. galaxies in regions of the ${\cal
M}$-$R_e$ in which there are no SDSS ETGs: for instance, three model 
galaxies with $\log {\cal M}/ M_{\odot}\sim 12$ and $R_e> 70$ kpc, 
and a model galaxy with $\log {\cal M}/ M_{\odot}\sim
11.5$ and $R_e\sim 0.3$ kpc (Fig. 3). We recall here that the existence
of compact low-$z$ ETGs with sizes and masses comparable to those of 
compact ETGs at $z>2$ is somehow unclear, with some results showing
an absence of such galaxies (e.g. Taylor et al. 2010) and others
finding a few candidates (e.g. Shih \& Stockton 2011; Valentinuzzi et 
al. 2010).
Do our results necessarily imply that dry merging alone cannot
explain the observed size evolution ? In principle it could be the case
that dry mergers are responsible for the whole size evolution, but the
actual rate of mergers is higher than predicted. This hypothesis can
be tested by further comparison with observations.  A first constraint
comes from the observed redshift evolution of the ETG stellar mass
function. Let us take, for instance, the model describing the
evolution of the $z \sim 1.7$ ETGs.  At each $z\leq 1.7$ we select
only model galaxies with $\log {\cal M}/ M_{\odot}\geq10.9$ and for
this subsample we measure the average stellar mass $\langle \log {\cal
  M}\rangle$. The redshift variation of $\langle \log {\cal M}\rangle$
is found to be well represented by the fit $\langle \log {\cal
  M}/M_{\odot}\rangle(z)=11.57-0.15z$, i.e. the average mass increase
by $\sim 30\%(70\%)$ from $z=0.7(1.5)$ to the present. This is
compatible with observed evolution of the ETG stellar mass function at
$0<z<1$ (e.g. Pozzetti et al. 2010).  Another testable feature of the
model is the predicted merger rate.  Let us define major(minor)
mergers those with mass ratio$>$($<$)1/4.  The predicted number of
major mergers per unit time $d N_m/d t$ decreases for decreasing $z$
(see Fakhouri et al. 2010): for our model ETGs we get, on average, $d
N_m/d t\sim 0.13(0.23)\,{\rm Gyr}^{-1}$ at $z\sim0.55(1.15)$. These
rates are higher by a factor of $\sim 2$ than estimated
observationally at similar $z$ by Bundy et al. (2009; see also Lotz 
et al. 2011), indicating that the considered model might be extreme also
in this respect. In the model, both major and minor mergers contribute
significantly to the growth of stellar mass (for instance, $\sim 50\%$ 
each between $z=1.5$ and $z=0$), so, at least within
$\Lambda$CDM, massive ETGs do not accrete most of their mass in very
minor mergers, which would be more effective in increasing the galaxy
size. The above arguments suggest that other processes not included in
the model should contribute significantly to the size
evolution. Unfortunately, at the moment the proposals for additional
mechanisms are not very promising: Fan et al. (2008) envisaged that
feedback from QSOs could play a role, but also this scenario is not
without problems (Ragone-Figueroa \& Granato 2011).

%%%%%%%%%%%%%%%%%%%% FIG 3
\begin{figure}
\includegraphics[width=0.98\linewidth]{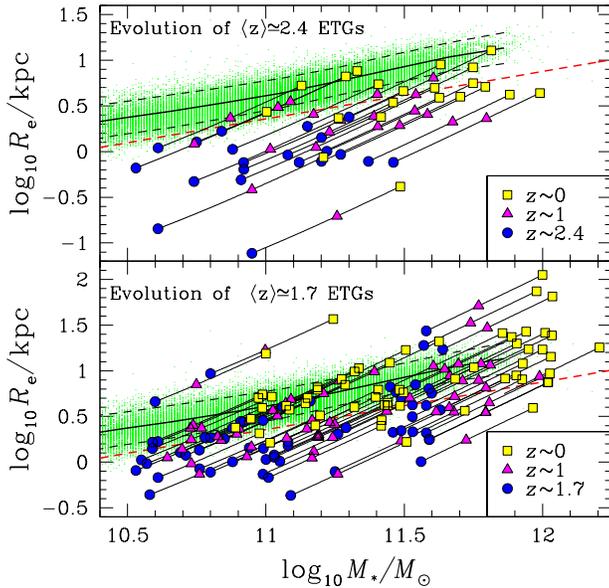}
\caption{Redshift evolution of ETGs in the stellar mass-effective
  radius plane, as predicted by the {\it dry}-merger $\Lambda$CDM
  model described in Section 5.  The progenitors (circles) are the
  ETGs observed at $2<z<3$ ($\langle z\rangle\simeq 2.4$, upper panel)
  and those observed at $1.5<z<2$ ($\langle z\rangle\simeq 1.7$, lower
  panel). The predicted descendants at $z=1$ and $z=0.14$ are
  represented, respectively, by the triangles and the squares. For
  comparison we plot SDSS ETGs (green dots) and the local SDSS
  best-fit with its observed scatter (black solid and dashed curves;
  Shen et al. 2003). The red dashed lines indicate the
  best-fits for ETGs observed at $0.8<z<1.2$.}
\end{figure}

\section{Conclusions}

The analysis of a large sample of ETGs at $0<z<3$ shows that their
size evolves independently of stellar mass and possibily faster at
$z>2$ (especially for ETGs with the largest masses). The interpretation
of this result is not straightforward as the available information
does not allow us to assess if this is an observational bias
(e.g. large ETGs with low surface brightness are missed in high-$z$
samples), or it is an intrinsic change in the evolutionary pattern
implying a very rapid growth of ETGs from $z>2$ to lower redshifts. We
explored the possibility of pure {\it dry} merging as the dominant
growth mechanism within the $\Lambda$CDM framework, and found that
this scenario is marginally consistent with the average size
evolution at $0<z<1.7$, but predicts descendants too compact for
$z>2$ progenitor ETGs.  Further studies and larger samples of ETGs
at $z>1.5$, which will be obtained with future wide-field surveys
(e.g. {\it Euclid}, Laureijs et al. 2011), will shed light on
these open questions.

\section{Acknowledgements}

We thank M. Bernardi, M. Moresco and V. Strazzullo for providing 
their data, J. Pforr and C. Maraston for providing their scaling 
relations, E. Daddi and A. Renzini for useful discussion, the 
COSMOS/zCOSMOS teams for making the data available to the community, 
and the anonymous referee for the constructive comments. AC is 
supported by grants ASI-Uni. Bologna-Astronomy Dept. I/039/10/0 
and PRIN MIUR 2008. CN is supported by grant PRIN MIUR 2008.
Paolo Cassata aknowledges support from ERC grant ERC-2010-AdG-26107-EARLY.

\label{lastpage}

\begin{thebibliography}{99}
\bibitem[]{}Behroozi P.~S., Conroy C., Wechsler R.~H., 2010, ApJ, 717, 379
\bibitem[]{}Bournaud F., Jog C.J. \& Combes F. 2007, A\&A, 476, 1179
\bibitem[]{}Boylan-Kolchin M., Springel V., White S.~D.~M., Jenkins A., 
Lemson G., 2009, MNRAS, 398, 1150
\bibitem[]{}Brammer G.B. et al. 2011, ApJ, 739, 24
\bibitem[]{}Buitrago F., Trujillo I., Conselice C., Bouwens R.J., 
Dickinson M., Yan H. 2008, ApJ, 687, L61
\bibitem[]{}Bundy K., Fukugita M., Ellis R.~S., Targett T.~A., Belli S., Kodama T., 2009, ApJ, 697, 1369
\bibitem[]{}Cappellari M. et al. 2009, ApJ, 704, L34
\bibitem[]{}Carrasco E.R, Conselice C.J., Trujillo I. 2010, MNRAS, 405,
2253
\bibitem[]{}Cassata P. et al. 2008, A\&A, 483, L39 
\bibitem[]{}Cassata P. et al. 2011, ApJ, 743, 96 
\bibitem[]{}Cenarro A.J., Trujillo I. 2009, ApJ, 696, L43 
\bibitem[]{}Cimatti A. et al. 2004, Nature, 430, 184
\bibitem[]{}Cimatti, A. et al. 2008, A\&A, 482, 21
\bibitem[]{}Cooper M.C. et al. 2011, ApJ, in press (arXiv:1109.5698)
\bibitem[]{}Daddi, E. et al. 2005, ApJ, 626, 680
\bibitem[]{}Damjanov I. et al. 2011, ApJ, 739, L44 
\bibitem[]{}di Serego Alighieri S. et al. 2005, A\&A, 442, 125
\bibitem[]{}Domi'nguez Sa'nchez, H. et al. 2011, MNRAS, 417, 900
\bibitem[]{}Fakhouri O., Ma C.-P., Boylan-Kolchin M., 2010, MNRAS, 406, 2267 
\bibitem[]{}Fan L., Lapi A., De Zotti G., Danese, L. 2008, ApJ, 689, L101
\bibitem[]{}Fontana, A. et al. 2009, A\&A, 501, 15
\bibitem[]{}Hyde J.B., Bernardi M. 2009, MNRAS, 394, 1978
\bibitem[]{}Hopkins P.F., Bundy K., Murray N., Quataert E., Lauer T.R., 
Ma C.-P. 2009, MNRAS, 398, 898
\bibitem[]{}Khochfar S. \& Silk J. 2006, MNRAS, 370, 902
\bibitem[]{}Kriek, M. et al. 2006, ApJ, L71
\bibitem[]{}Laureijs R. et al. 2011, Euclid Definition Study Report (arXiv:1110.3193)
\bibitem[]{}Lotz J.~M., Jonsson P., Cox T.~J., Croton D., Primack
  J.~R., Somerville R.~S., Stewart K., 2011, ApJ, in press (arXiv:1108.2508)
\bibitem[]{}Mancini, C.; Daddi, E.; Renzini, A. et al. 2010,
MNRAS, 401, 933
\bibitem[]{}Maraston C. 2005, MNRAS, 362, 799
\bibitem[]{}Moresco M. et al. 2010, A\&A, 524, 67 
\bibitem[]{}Naab T., Johansson P.H., Ostriker J.P. 2009, ApJ, 699, L178
\bibitem[]{}Newman A.B., Ellis R.S.; Bundy K., Treu T. 2011, ApJ, in press 
(arXiv:1110.1637)
\bibitem[]{}Nipoti C., Londrillo P., Ciotti L., 2003, MNRAS, 342, 501
\bibitem[]{}Nipoti C., Treu T., Bolton A.~S., 2009a, ApJ, 703, 1531
\bibitem[]{}Nipoti C., Treu T., Auger M.W., Bolton A.S. 2009b, ApJ, 706, L86
\bibitem[]{}Nipoti C., Treu T., Leauthaud A., Bundy K., Newman A.~B., Auger M.~W., 2012, MNRAS in press (arXiv:1202.0971)
\bibitem[]{}Onodera M. et al. 2010, ApJ, 715, L60 
\bibitem[]{oser} Oser L., Naab T., Ostriker J.P., Johansson P.H. 2011, ApJ, 
in press (arXiv:1106.5490) 
\bibitem[]{}Papovich C. et al. 2011, ApJ, in press (arXiv:1110.3794)
\bibitem[]{}Pforr J., Maraston C., Tonini C. 2012, MNRAS, submitted 
\bibitem[]{}Pozzetti, L. et al. 2010, A\&A, 523, 13
\bibitem[]{}Ragone-Figueroa C., Granato G.~L., 2011, MNRAS, 414, 3690
\bibitem[]{}Renzini A., 2006, ARA\&A, 44, 141 
\bibitem[]{}Rettura A. et al. 2010, ApJ, 709, 512
\bibitem[]{}Scarlata C. et al. 2007, ApJS, 172, 406
\bibitem[]{}Saglia R.P. et al. 2010, A\&A, 524, 6
\bibitem[]{}Saracco P., Longhetti, M., Andreon, S. 2009, MNRAS, 392, 718
\bibitem[]{}Shen S. et al. 2003, MNRAS, 343, 978
\bibitem[]{}Shih H.-Y., Stockton A. 2011, ApJ, 733, 45 
\bibitem[]{}Springel V., et al., 2005, Nature, 435, 629
\bibitem[]{}Strazzullo V. et al. 2010, A\&A, 524, 17
\bibitem[]{}Taylor E.N., Franx M., Glazebrook K., Brinchmann J., van der Wel A., van Dokkum P.G. 2010, ApJ, 720, 723
\bibitem[]{}Trujillo I. et al. 2006, MNRAS, 373, L36 2006, MNRAS, 373, L36
\bibitem[]{}Valentinuzzi T. et al. 2010, ApJ, 721, L19 
\bibitem[]{}van de Sande J. et al. 2011, ApJ, 736, L9 
\bibitem[]{}van der Wel A. et al. 2008, ApJ, 688, 48 
\bibitem[]{}van der Wel A. et al. 2011, ApJ, 730, 38 
\bibitem[]{}van Dokkum P.G. et al. 2008, ApJ, 677, L5
\bibitem[]{}van Dokkum P.G., Kriek M., Franx, M. 2009, Nature, 460, 717 
\bibitem[]{}van Dokkum P.G., Brammer G. 2010, ApJ, 718, L73 
\bibitem[]{}Williams R.J., Quadri R.F., Franx M., van Dokkum P., 
Toft S., Kriek M., Labbe´ I. 2010, 713, 738
\end{thebibliography}
\end{document}